# Scheduling in Grid Computing Environment

Harshadkumar B. Prajapati
Information Technology Department
Dharmsinh Desai University
Nadiad, INDIA
e-mail: prajapatihb.it@ddu.ac.in,
harshad.b.prajapati@gmail.com

Vipul A. Shah
Instrumentation & Control Engineering Department
Dharmsinh Desai University
Nadiad, INDIA
e-mail: vashah.ic@ddu.ac.in

*Abstract*—Scheduling in Grid computing has been active area of research since its beginning. However, beginners find very difficult to understand related concepts due to a large learning curve of Grid computing. Thus, there is a need of concise understanding of scheduling in Grid computing area. This paper strives to present concise understanding of scheduling and related understanding of Grid computing system. The paper describes overall picture of Grid computing and discusses important sub-systems that enable Grid computing possible. Moreover, the paper also discusses concepts of resource scheduling and application scheduling and also presents classification of scheduling algorithms. Furthermore, the paper also presents methodology used for evaluating scheduling algorithms including both real system and simulation based approaches. The presented work on scheduling in Grid containing concise understandings of scheduling system, scheduling algorithm, and scheduling methodology would be very useful to users and researchers.

I. INTRODUCTION

Research and development in Grid computing [1] started with the aim of using free or idle resources that are geographically distributed, administratively decentralized, and heterogeneous in capability and speed for performance demanding scientific applications. Grid computing has been used by many research projects and research in Grid has reached to many places in academia. However, due to a long learning curve of Grid computing, many interested researchers stay away from it due to substantial time needed in understanding a big literature work on Grid computing. Before research work on scheduling [2] in Grid computing is attempted by researchers or beginners, the researchers must be familiar with important concepts of systems and computer networking. Without certain prerequisite understanding of networking fundamentals, protocols, process control and management, research in Grid computing can not be thought of. The objective of this research paper is to provide concise understanding of Grid scheduling area with discussion of important concepts and applied tools/software.

Scheduling aspect is found in operating system, which manages execution of a process at various stages of the life-cycle of the process. Scheduling of processes in OS is handled by long-term, medium-term, and short-term schedulers. CPU scheduler in OS allocates CPU(s) to the live processes. It is necessary to understand how scheduling in OS is different from scheduling in cluster computing and Grid computing. Moreover, it is also necessary to understand why scheduling algorithms that are available in OS cannot be used for scheduling jobs of Grid computing. Beginners need to spend a lot of time in understanding various topics related to Grid computing. We take *scheduling in Grid* as a topic of research and try to provide concise understanding of all important concepts pertaining to scheduling of jobs on Grid computing infrastructure. Main objective of this paper is to provide various understandings related to *scheduling in Grid* at a single space, i.e., in this paper. Scheduling concepts in a Grid system is present at the different layers including OS, cluster, and global system. It is necessary to understand objective of each, reason for presence of each, and overall understanding.

Though the literature on Grid computing covering exhaustive details and understandings on workflow management system (WMS), resource management system, and scheduling algorithms exist, beginners find very difficult to understand various concepts and complex systems unless they are presented with a big-picture in context. The work in [3] presents various workflow management systems based on various selected criteria. It provides comprehensive evaluation of various WMSs. The work in [4] presents understanding of various resource management systems (RMSs) that can be used for utility driven cluster computing. The work in [5] presents an abstract/generalized model of RMS. The work provides taxonomy and understanding of various concepts related to RMS. The work discusses concept of scheduling with internal organization of scheduler, state estimation of resources, and about scheduling policy that does ordering of requests or jobs. We redirect readers to [5] for getting details on various Grid resource management systems. Various concepts and understandings are available in the literature; however, they are available at scattered locations. This paper strives to provide overall picture of scheduling in Grid at a single place, i.e., in this paper, and presents concepts related to Grid computing in a magnifying view.

This paper is structured as follows. Section II discusses distributed computing environments in current use, presents fundamental concepts of scheduling, and relate scheduling in Grid with scheduling in *Systems*. Section III concisely discusses various sub-systems or systems related to Grid computing and also presents discussion on Grid simulation tools. Section IV covers scheduling algorithms related to Grid computing falling into two main types: resource scheduling and application scheduling. Section V highlights methodology of evaluation of scheduling algorithms using both real system and simulation tool based approaches. Finally, Section VI provides conclusion.

## II. IMPORTANT CONCEPTS RELATED TO SCHEDULING IN GRID COMPUTING ENVIRONMENT

It is very important to understand difference between Grid computing environment, which is of type distributed computing, from other distributed computing environments. We first like to present brief introduction about types of distributed computing environments that are widely used for solving problems in distributed manner in the current era of network-computing.

### A. Currently used Distributed Environments

*1) Cluster Computing*

Cluster is a connected group of loosely coupled computing machines. Cluster computing [6] uses master-slave based centralized scheduling architecture. Jobs are submitted to master. Master schedules it on slave. In cluster, all slaves nodes are generally homogeneous having uniform CPU speed, memory, and network bandwidth. To build a Cluster computing environment, a cluster middleware software need to be installed and configured on network of LAN computers. Main goal of Cluster computing is better system performance, which is handled by a central resource manager.

*2) Grid Computing*

Grid computing enables utilization of idle time of resources that are available at geographically diverse locations [7], [8], [9]. Grid computing can allow access of resources such as storage, sensors, application software/code, databases, and computing power. In Grid, resources are autonomous and heterogeneous. Current Grid computing, in most deployments, is on collaborative manner, in which resources exhibit varying availability. However, QoS oriented Grid computing is also possible. Grid computing has been used in drug discovery, GIS processing, sky image processing, industrial research, and scientific experiments. Use of Grid computing infrastructure is done by converting traditional applications into Grid applications. The independent tasks applications include Parameter Sweep and Task farming (embarrassingly parallel) problems [10], e.g. drug discovery. Dependent tasks applications include workflow [11] applications, e.g., Montage [12].

Figure 1 shows the constituent entities that make Grid computing environment, a large Virtual Organization [13] formed of many execution Grid sites owned by different real organizations, e.g., Grid site 1, . . ., Grid site M, for execution of Grid applications, independent tasks or dependent tasks. In Grid, resources are usually autonomous and the Grid scheduler does not have full control of the resources, i.e., Grid schedulers cannot violate local policies of resources and usually there might not be any central Grid scheduler. Generally, each Grid site can have its own Grid scheduler and same local resource can be part of more than one VO. Resource scheduler at each Grid site schedules and executes individual jobs irrespective of from which Grid applications they came from. Grid scheduler, which is also called global scheduler or meta-scheduler or super scheduler or broker, schedules jobs of Grid applications on available Grid sites with focus on satisfying application specific scheduling objective(s). Availability of resource monitoring infrastructure at each participating Grid site can provide substantial guidance to Grid scheduler for making scheduling decision. Job monitoring provides monitoring of status of execution of jobs. Having appropriate system components, Grid computing can enable use of resources based on availability, capability, performance, cost and users quality of service requirement.

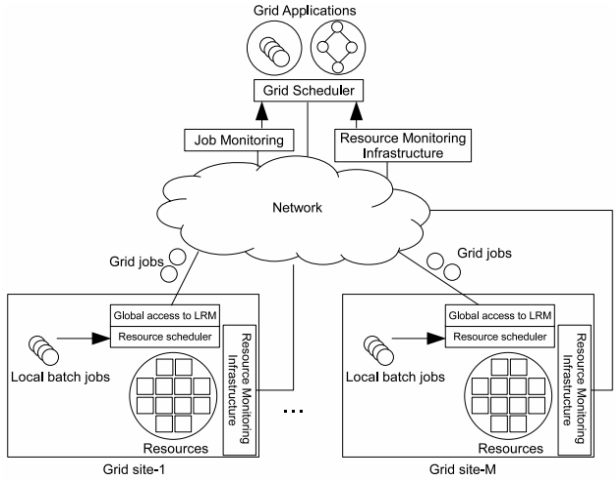

Figure 1. An architectural diagram showing core entities of Grid computing.

*3) Cloud Computing*

Cloud computing [14] exploits virtualization and Internet technologies for providing needed resources, hardware or software, on demand and on lease to resource consumers. Thus, Cloud computing can facilitate e-Commerce business without spending money on computer-system infrastructure by transferring worries of infrastructure and software to Cloud services provider. Using Cloud computing, hardware, software, etc. can be bought as services and consumers need to pay only for what and how much they use. Architecture of Cloud computing includes data-center made of SAN or NAS, virtualization server, racks of physical processors, high speed network, and virtual machines management software. Important research issues in Cloud computing involve imposing, monitoring, and management of Service Level Agreements and scheduling and allocation of virtualized resources on demand.

B. *Scheduling Fundamentals and Terminology*

*1) What is Scheduling?*

Scheduling is a decision process of assignment of the relatively large number of jobs to the relatively small number of workers in either space or time or both dimensions in order to achieve desired objective(s). This decision may depend on duration of tasks/activities, predecessor tasks/activities, resource availability, and target completion time depending upon target environment made of workers, types of jobs, and desired requirements pertaining to the completion of jobs. For example, scheduling in Operating System allocates CPU, *a worker*, to the live processes, *jobs*, running in the system. Scheduling is also used in multi-processor systems, automated manufacturing systems, flight scheduling, train scheduling, etc.

*2) Scheduling Terminology*

- **Release-time (date)**: The earliest time at which a job can start its processing.
- **Processing-time**: The time-duration needed by a job to finish its processing.
- **Start-time**: The actual time at which a job starts its processing.
- **Finish-time**: The actual time at which a job finishes its processing.
- **Expected execution time**: It is a time expected to be taken up by a job to finish its execution. It is derived empirically.

C. *Scheduling in Systems*

*1) Scheduling in Operating System v/s Scheduling in Grid*

If scheduling decision is taken by OS, then why one needs to learn one more scheduling, i.e., Grid Scheduling. In Operating System, most applications are user interactive. OS does instruction level (CPU) scheduling. It exploits interleaving of I/O operations and CPU operations. The main objective of scheduling in OS is fairness among processes. In Grid environment, most applications are not user interactive. Grid scheduling works at task level in which parallelism of independent tasks is exploited. The main objective in Grid scheduling is finish job or application at earliest. The scheduling of OS is of type local scheduling as it determines how the processes resident on a single CPU are allocated and executed. The scheduling in Grid is of type global scheduling as it allocates processes to multiple resources to optimize system wide performance objective or application specific objective.

*2) Relating Scheduling in Grid with Multiprocessor Scheduling*

There exists an α, β, γ classification scheme [15] for classification of scheduling problem. The α, β, γ classification scheme of scheduling problems was originally presented by Graham et al. and later was extended by Veltman et al. Although it is for multiprocessor environment, it can also be used for representing problem of scheduling on distributed environment and the classification scheme has been used by many researchers in their work. TABLE I compares scheduling in multiprocessing system [16] and in Grid computing system.

TABLE I. RELATING SCHEDULING IN GRID SYSTEM WITH SCHEDULING IN MULTIPROCESSOR SYSTEM

| Scheme | Multi-processor Environment | Grid Environment |
|---|---|---|
| α: Processor Environment | **Processor type**: identical processors, heterogeneous processors | **Grid Resources**: Compute resources with different LRMs, Data resources, Network resources, Sensor resources |
| | Number of processors | Number of Grid Resources |
| β: Task Characteristics | Independent or have precedence constraints | Independent jobs or workflow |
| | Computation costs | Computation costs |
| | Communication costs | Communication costs |
| | Task duplication allowed | Task duplication allowed |
| γ: Optimality Criteria | Completion time of a task | Completion time of a Grid application |

*3) Relating Machine Scheduling with Grid Scheduling*

The machine scheduling problem is mentioned here to make reader familiar with how Grid scheduling and machine scheduling [2] are related. The novel approaches used for machine scheduling problem can be explored for their applicability in Grid scheduling. In machine scheduling problem, there are $m$ machines, $n$ jobs, and $m$ tasks (operation in each job) per job. Three shop scheduling problems are concisely presented below.

- **Open-shop scheduling**: Different tasks of the same job may be scheduled in any order. If preemption of tasks is allowed, the tasks can be interleaved.
- **Flow-shop scheduling**: If the order of tasks is fixed and it is same for all jobs, then scheduling is called flow-shop scheduling.
- **Job-shop scheduling**: In this scheduling, tasks of a job are totally ordered, But the order of tasks is possibly different for each job.

Based on above definition of different machine scheduling problems, the problem of scheduling of independent tasks in Grid can be related to the Open-shop scheduling and the problem of scheduling of dependent tasks in Grid can be related to the Job-shop scheduling. Scheduling of looping part of the control flow graph can be related to Flow-shop scheduling.

*D. Characteristic of the problem of scheduling in Grid*

In Grid computing, the problem of scheduling involves a number of tasks and a number of resources. General objective of scheduling in Grid is to generate optimal mapping of the tasks onto the resources. This mapping of tasks in distributed environment is NP-Complete [17] problem. Therefore, we need to use heuristic to solve such problems in polynomial time. A workflow scheduling is a specialized version of scheduling of jobs in distributed environment, in which there exists dependencies among jobs.

III. SYSTEMS FOR GRID COMPUTING AND SCHEDULING

This section presents understandings of various sub-systems/systems that could be used in implementing Grid computing with scheduling capability. Moreover, the section also focuses on simulation based study in Grid computing.

*A. Grid System*

*1) Grid Middleware*

Grid middleware is a software that glues diverse local resources of various organizations to form a higher-level, bigger, global resource. Grid middleware offers following services: (1) remote process/job management to submit and to monitor, (2) allocation and co-allocation of resources, (3) access to storage devices and data management, (4) information service (kind of yellow page service) including resource discovery, resource registration, and resource information update, and (5) security service including authentication, authorization, delegation, and single-sign on. Example of Grid middleware are as follows: Globus [8], Legion [9], and Unicore [7]. Globus is a de facto standard software for building Grid computing and it has been widely used in many Grid deployments.

*2) Batch Queue Controlled Resources*

In most Grid deployments, individual resources of a single organization are centrally managed using batch queue oriented local resource management system (LRM). In such system, user submitted batch jobs are introduced into the queue of LRM, from which resource scheduler decides about execution of jobs, see Section IV,A. Examples of such LRMs include PBS [18], LSF [19], Condor [20], Sun Grid Engine [21], NQE [22], Maui [23] and Load Leveler [24]. A resource scheduler is also known as low-level scheduler or local scheduler. For compute resources, LRM can be configured for (i) which user is allowed to run jobs (ii) what policies are associated with selection of jobs for running, and (iii) what policies are associated with individual machines for considering them to be idle.

A LRM manages two types of jobs: local jobs generated inside a resource domain and jobs generated by external users, i.e. Grid users, see Figure 1. The common purpose of any LRM is to manage, control, and schedule batch processes on the resources under its control. Since, LRM deals with distributed resources, it is also called Distributed Resource

Manager. Basic features of any LRM include following: (1) scripting language for defining batch job, (2) interfaces for submission of batch jobs, (3) interfaces to monitor the executions, (4) interfaces to submit input and gather output data, (5) mechanism for defining priorities for jobs, (6) match-making of jobs with resources, and (7) scheduling jobs present in queue(s) to determine execution order of jobs based on job-priority, resource status, and resource allocation configuration. Advance reservation based LRMs are also used in Grid. Examples of such LRMs include Platform LSF, PBS Pro/Torque, Maui, and SGE.

*3) Connecting Resources of different organizations into a VO*

Grid allows use of heterogeneous LRMs through a common interface. In Globus based Grid, which is widely used, GRAM (Globus Resource Allocation Manager) interface is used for job submission on heterogeneous LRMs. GRAM is a standardized way of accessing any LRM, as it is LRM neutral. GRAM messages remain same irrespective of LRM, whether it is PBS, LSF, or Condor. Each LRM can be connected into a Globus based Grid using Globus-LRM adapter. Clients submit job requests using GRAM protocol and GRAM-LRM connector can translate those messages into a language understood by a specific LRM. For example, GRAM5 supports Condor, PBS, and LSF; similarly, Unicore supports SGE, LoadLeveler and Torque.

*B. Resource Monitoring Infrastructure*

Resource monitoring helps to Grid scheduler in taking appropriate decision on available resources. Many resource monitoring frameworks are available. We discuss two widely used resource monitoring systems: one provides information on individual resources and another provides information on connection between resources. Network Weather Service (NWS) and Ganglia are most widely used network performance and resource monitoring systems in real Grid or cluster test-beds.

*1) Ganglia*

Ganglia monitoring system [25] provides information on individual resources of Clusters or Grids. Ganglia exploits scalability, redundancy, and decentralization features of distributed computing system. Ganglia system has hierarchical design, which is made of *gmond* and *gmetad* component sub-systems. The *gmond* (Ganglia monitoring daemon), which runs on each node of a cluster, enables monitoring on a single cluster through listen/announce protocol. A collection of hierarchical *gmetad*s federate multiple clusters through a tree of TCP connections to prepare aggregated view. Ganglia uses XML to represent client's response, XDR for quick data transport and efficient storage, and *RRDtool* for round robin data storage and visualization of time series data. Ganglia provides command line access through gmetric command and programming access through client side library. The *gmond* supports two types of metrics: (1) 28 to 37 standard/in-built metrics, depending upon operating system, and (2) user-defined metrics. Ganglia has been used in clusters, Grids, and Planetary-scale systems. Ganglia also provides PHP based web front-end, which enables various views of monitoring data rendered through *RRDtool*.

*2) Network Weather Service*

Network Weather Service (NWS) [26] provides dynamic characteristics of networked computing resources. Moreover, it can also forecast [27] performance of resources. Resource monitoring and performance prediction provided by NWS is useful to higher level services to enable scheduling and QoS guarantees in Grid or cluster computing environments. Its implementation is based on Unix/Linux platform using TCP/IP socket based communication among distributed processes. NWS is made of four component systems: (1) Persistent State process, (2) Name Server process, (3) Sensor process, and (4) Forecaster process. The Persistent State process stores measurements in persistent storage. It also provides access of measurements to their users. NWS supports measurements of following variables: fraction of CPU availability time, connection time of TCP socket, end-to-end network latency of TCP, and end-to-end network bandwidth of TCP. NWS implements CPU availability sensor as passive sensor, which itself does not measure any characteristics, rather relies on *vmstat* utility. The network sensor is implemented as active sensor, which explicitly measures network performance related measures by using active measurement probes.

*C. Application Scheduling Systems*

We discuss two types of scheduling systems: scheduling system for independent tasks applications and scheduling system for dependent tasks applications.

*1) Independent Tasks Scheduling Systems*

Scheduling of application, whether independent tasks or dependent tasks, in Grid computing involves following steps: resource discovery, resource selection, schedule generation, and application execution. In case of dynamic scheduling, the process may involve additional step of rescheduling or schedule adaptation. General architecture of Grid supporting scheduling of application is discussed in Section II,A,2. Examples of systems supporting scheduling of independent tasks applications are as follows: Nimrod-g [10], AppLeS [28], NetSolve/GridSolve [29] and ICENI [30].

*2) Dependent Tasks Scheduling System---Workflow Management Systems*

A workflow management system (WMS) is used to schedule dependent tasks applications. A workflow contains a sequence of tasks that are to be carried out in the defined order in order to generate desired end-result. There are two common types of workflows: Business Workflow and Scientific Workflow. The business workflow is related to business application. It automates a business process fully or partially. Some business workflow may include human decision to make further processing steps of the workflow. In Business workflow, documents, information, and business decision are passed from one participant to another. Scientific workflow automates executing computation tasks of scientific application. It involves analytical steps. The analytical steps may involve one or more of following: database access and querying steps, mathematical processing steps, data analysis and mining steps, and many other steps that involve any computationally intensive activities. Workflow model is graphical representation of the workflow steps. Two types of workflow are mainly used: Data flow (DAG based) and Control flow (Petri-net) based. DAG based is mainly used in current Scientific Workflow in which loops and control structures by definition are not available. Control flow (Petri-net based) is mainly used in business workflows. Workflow language bridges the gap between graphical workflow model and the workflow engine, which executes the workflow. Workflow engine understands workflow language. A workflow can be written directly in the workflow language without preparing workflow model by GUI. Most workflow languages are XML based for both scientific and business workflows.For representation of workflow.

Various Workflow Management Systems (WMSs) support either Directed Acyclic Graph based workflow or Control Flow Graph based workflow or both. Triana [31], GridAnt [32] or Karajan [33], UNICORE [7], Askalon [34], and ICENI [35] are Control Flow Graph based WMSs and DAGMan [36], Taverna [37] , GrADS [38] GridFlow [39], Gridbus [40] , and Pegasus WMS [41] are DAG based WMSs. Most of the mentioned systems are Globus based except Taverna, GridFlow, and UNICORE. Furthermore, a few systems also support web-services. Only a few systems are active in further development of WMSs and providing help or support to their users. We redirect readers elsewhere, for example, to [3], for understanding of various WMSs.

*D. Grid Simulation Tools*

Scheduling algorithms in Grid can be evaluated using Grid simulation tools also. However, accuracy of simulation results depends on how accurately simulator mimics real system entities in simulation. There is another approach of performance evaluation called mathematical analysis; however, for beginners it is not advisable unless they have sound understanding of mathematical representation and queuing based mathematical analysis [42].

There exists many simulation tools or simulation frameworks that support simulation based study in Grid environment. However, they differ based on kind of studies supported by them. We briefly mention each Grid simulation tool with supported study details. Following are various open source simulation tools. SimGrid [43] is a C language based generalized toolkit supporting DAG applications and MPI applications. GridSim [44] is a Java language based generalized framework supporting various entities of Grid computing. OptorSim [45] is Java language based simulator and it can simulate compute Grid and data Grid [46]. It focuses on data replication strategies and replica optimization. Alea [47] is Java language based simulator. It supports study of various resource (Local) scheduling algorithms. GSSIM [48] is Java language based framework. It can support concept of scheduling at local/resource level and Grid level. Moreover, GSSIM can understand existing workloads generated using real Grid environment. It also supports generating synthetic workload. MaGate [49] is a Java language based simulator that supports peer-to-peer (P2P) Grid. Study using MaGate includes resource discovery algorithms in P2P Grid and bio-inspired community scheduling algorithms.

IV. CLASSIFICATION OF THE SCHEDULING USED IN GRID ENVIRONMENT

Two main types of scheduling are involved in Grid, one at local resource level and another at application level. We concisely explore them in following discussion.

*A. Resource Scheduling*

Two types of scheduling: time sharing and space sharing are involved at local resource level. Time sharing scheduling is used by each machine (computer) of the cluster, for example, CPU scheduler of the Operating System on each machine that is part of batch system, whereas space sharing scheduling is used by Local Resource Manager to schedule the job present in batch queue on idle machine of the cluster. Resource scheduling is used to increase utilization of resources or balance load on the resources. Generally a Grid site contains a batch queue controlled cluster, see Figure 1. Resource scheduling is not complicated as it considers each submitted task an independent unit and therefore the scheduler can mix in with or order with tasks of other applications, either submitted from the same Grid site or from other Grid sites.

The resource scheduling involves taking two decisions: job selection from a queue/bag of jobs and node selection for selected job from available nodes. Figure 2 provides classification of algorithms under the two main decision operations, which execute as part of resource scheduling.

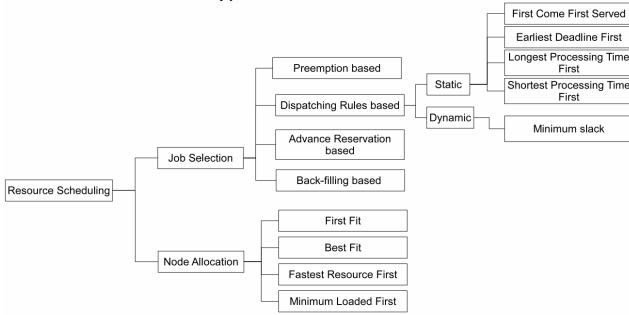

Figure 2.  Classification of Resource Scheduling in Grid Computing.

**Job Selection Algorithms**:

Job selection algorithm selects a job from a queue of jobs. The simplest type of job selection algorithms are *Dispatching Rules based*. FCFS [51] schedules jobs as per submission order (time). *Shortest Processing Time First* or Short-Job-First (SJF) assigns highest priority to the job that has shortest processing time. *Longest Processing Time First* or Largest-Size-First (LSF) assigns highest priority to the job that has longest processing time. *Earliest Deadline First* (EDF) assigns highest priority to the process that has expected execution time in the most near future or the specified deadline is in the most near future. In dynamic dispatching rules based scheduling, decision changes depending upon passage of time. Minimum slack based algorithm is of type dynamic dispatching rules based scheduling. In this algorithms, jobs are ordered according to their remaining slack, which is defined as $slack = max(d_j - p_j - t, 0)$, where $d_j$ is deadline, $p_j$ is processing time, and $t$ is the current time. Backfilling is an optimization of FCFS algorithm. If the oldest job in the queue cannot be executed due to non-availability of enough/required resources, the older jobs are started without delaying the oldest queued job though the older job arrived latter than oldest job. Condor [52], LSF [19], and PBS [18] supports backfilling based resource scheduling algorithm. In Advance Reservation based Algorithm [51], user of resource sends reservation-request for using resource in future time. This advance reservation based algorithm can guarantee that required resource will be available to user for use; moreover, jobs do not need to wait in queue. Examples of systems supporting advance reservation based algorithms are as follows: Platform LSF, PBS Pro/Torque, Maui, and SGE. Preemption based resource scheduling algorithm can stop execution of low priority job when high priority job arrives.

**Node Allocation Algorithms**:

*Best Fit* selects the node that has the fewest available resources and still can finish the job. *First Fit* schedules the job on the first suitable resource in the list of available resources. *Fastest Resource First* selects fastest available resource in the list of available resources. *Min Loaded First* selects the resource having maximum unused CPU power or CPU utilization.

B. *Application Scheduling*

Grid scheduler, to which users submit their applications, performs application scheduling, in which the scheduler takes scheduling decision about tasks of whole application. Various terms are used for the term Grid scheduler. These terms include super-scheduler, meta-scheduler, global scheduler, application broker, and application scheduler. Important difference between resource scheduling and application scheduling is that application level scheduler assigns a Grid site to a job whereas resource scheduler allocates a machine of a cluster to a job. Figure 3 provides broad classification of application scheduling algorithms based on various criteria.

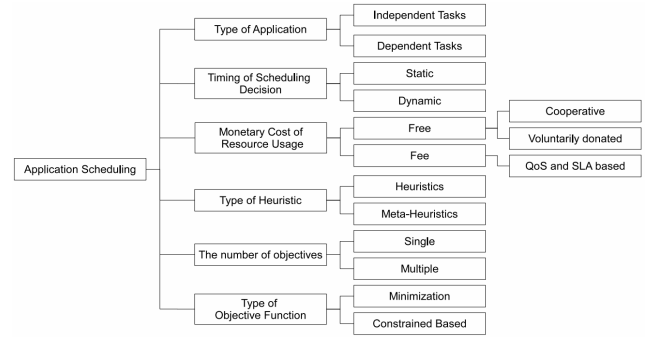

Figure 3.  Classification of Application Scheduling in Grid Computing.

Based on time at which scheduling decision is taken, scheduling can be categorized into two: Static scheduling and Dynamic scheduling.

**Static scheduling**: It determines schedule of all the tasks of application before application starts to run. Static scheduling is applicable to both independent tasks and dependent tasks applications. It is applicable when the environment is assumed to be static or non-changing and when it is possible to estimate the execution times of tasks (jobs).

**Dynamic scheduling** [53]: It determines schedule of a task of an application only when the task becomes ready for execution. For scheduling of independent tasks application, all tasks can be scheduled at the same time and in any order; however, for scheduling of dependent tasks application, a task becomes ready only when all its predecessor tasks have completed their execution. Dynamic scheduling is used when the environment is dynamic/changing. It is used when it is impossible to determine the execution time or when tasks (jobs) are coming online (real-time).

Exact algorithm always produces an optimal solution if exists. Heuristic produces a feasible and nearly optimal, but not necessarily optimal solution. The important difference between heuristic and meta-heuristic is mentioned below.

**Heuristics**: This approach develops a scheduling algorithm that fits a particular kind of problem. Examples of heuristics include following: Myopic, Batch scheduling, List scheduling, Clustering/grouping based scheduling, and Duplication based scheduling [54].

**Meta-heuristics**: This approach develops a general approach of developing a specific heuristic to fit a particular kind of problem, for example, Genetic Algorithm, Simulated Annealing, Tabu Search, and Particle Swarm Optimization [54].

Based on characteristics of application, scheduling can be categorized into independent tasks scheduling and dependent tasks/workflow scheduling.

**Independent tasks scheduling**: In independent tasks application, all tasks of applications are independent of each other in terms of any dependencies. Therefore they can be scheduled in any order by scheduler. Examples of independent tasks scheduling algorithms [55] include MET (Minimum Execution Time), MCT (Minimum Completion Time), Min-min, Max-min, and Suffrage [56].

**Dependent tasks scheduling**: In dependent tasks application, there exists dependencies among tasks of applications. Therefore the application scheduler has to respect these dependencies among tasks while scheduling them on resources. There are three types of heuristics [55]: list-scheduling, clustering/grouping based, and duplication based for scheduling of dependent tasks scheduling [55].

Whether resource has associated cost of usage or not, scheduling can be divided into following two categories.

**Cooperative/volunteer access based scheduling**: Scheduler tries to minimize the execution time ignoring other factors. This approach is currently used by most Grid projects present in the world. In this type of scheduling, the resource consumers can use resources freely.

**QoS and SLA based scheduling**: In this type of scheduling, which is also called utility based Grid computing [57], resource consumers need to pay to resource providers for usage of resources. Thus, monetary cost is also involved as decision factor for taking scheduling decision [58].

Based on how many objectives are involved in scheduling decision, scheduling can be divided into following two broad categories.

**Single objective scheduling**: In this scheduling, the scheduler has to focus on only one objective while taking scheduling decision. Example of scheduling objectives are time minimization and cost minimization.

**Multiple objective scheduling**: In this scheduling, the scheduler has to focus on multiple objective at the same time while taking scheduling decision. Example of multiple objective scheduling is cost and time minimization scheduling.

Based on characteristic of objective function, scheduling can be divided into following two broad categories.

**Minimization based scheduling**: In this scheduling, the scheduler tries to minimize objective. Objective could be minimization of make-span or minimization of cost.

**Constrained based scheduling**: In this scheduling, the scheduler is given information about upper-bound on the value of the objective and the scheduler has to generate schedule that satisfies objective value within the specified upper-bound.

## V. METHODOLOGY TO EVALUATE PERFORMANCE OF SCHEDULING ALGORITHMS

A proposed scheduling algorithm can be evaluated by one of two approaches: implement the proposed algorithm in a real scheduling system and evaluate for performance or implement the proposed algorithm in a simulator and evaluate for performance. Depending upon type of scheduling: resource or application, appropriate simulation tool or real system should be chosen. When a real system is to be built for evaluation of the proposed scheduling algorithms, it is very important to decide about what combination of available Grid middleware software and Grid application scheduler system can address the proposed algorithms. There are many options such as Globus, Condor-G, LSF, GridbusBroker, Pegasus, Kepler, Taverna, Triana, Grid(Lab) Resource Management, Nimrod-G, Sun Grid Engine, etc., see Section III. This section addresses methodology of carrying out evaluation of scheduling algorithms with focus on generalized steps, various types of test data, configuration of test environment, and performance evaluation metrics.

A. *Generalized steps of evaluation of scheduling algorithms*

Configuration of simulator is generally found to be easy; however, the exact way of configuration of a real system depends on selection of application scheduling software and grid middleware software. In following, we try to point out general steps of experimentation.

**Step 1**: Choose Grid system/WMS/simulation tool for carrying out scheduling.

**Step 2**: Choose test applications or batch-jobs or workflows or task-graphs.

**Step 3**: Choose scenarios for evaluation of algorithms.

**Step 4**: Choose performance metrics for comparison.

**Step 5**: (If needed,) Instrument system or tool for gathering data of interest for chosen metrics.

**Step 6**: Perform different experiments with manipulating independent variables. Include best, known (benchmark) scheduling algorithm for comparison with.

**Step 7**: Gather and analyze data or result through instrumented code or log records or trace records.

B. *Test data/input for evaluation of scheduling algorithms*

We focus on only testing of workflow applications.

Following are two main ways of generating test input data:

**Use well-known DAGs**: A few examples of well-known DAGs/workflows include EMAN (Electron Micrograph Analysis) BLAST (Basic Local Alignment Search Tool), Gaussian Elimination (Provides solution of a system of linear equations), Fast Fourier Transform, Montage, LIGO, and Galactic Plane.

**Prepare input data using DAG generator**: Use DAG generator to prepare random task graphs by varying following parameters: the total number of tasks, amount of communication, amount of computation, the number of parallel tasks, and degree of parallelism.

We need to configure resources/system to enable different test scenarios. It should be noted that in real system, it won't be possible to configure resources to achieve certain behavior. However, in simulation, we can vary the number of resources and characteristics of resources. In characteristics of resources, we can vary computation speed of resource, network bandwidth of resource, resource availability, and resource failure.

For exhaustive evaluation, different task graphs with known and unknown characteristics can be prepared and can be used in experimentation. A research paper on *Benchmarking the Task Graph Scheduling Algorithms* by Yu-Kwong Kwok and Ishfaq Ahmad in [59] provides details of benchmark task graphs. We concisely present them below, which researchers can choose from depending upon availability of task-graphs with them and feasibility of getting required ones with extra efforts.

**Peer set graph**: Use example task graphs used by various researchers and those are documented in publications.

**Random Graphs with optimal schedules**: Use task-graphs for which we have obtained optimal solutions using a branch-and-bound algorithm.

**Random Graphs with Pre-Determined Optimal Schedules**: This method generates graphs with known optimal schedules.

**Random Graphs without Optimal Schedules**: Use random graphs having large number of tasks (50 to 500) with varying value of size, communication-to-computation ratio (CCR), and parallelism.

**Traced Graphs**: Use task-graphs that are produced using a parallelizing compiler on some numerical parallel application programs.

C. *Performance evaluation metrics of scheduling algorithms*

This sub-section presents various performance evaluation metrics that are used for evaluation of resource scheduling algorithms and application scheduling algorithms. Performance evaluation metrics for resource scheduling algorithms are as follows.

**System utilization**: It indicates percentage of time the resource is busy.

**Throughput**: It indicates the number of jobs processed/finished in a given time period.

**Turnaround time**: It is also called response time of a job. It is defined as the sum of waiting time and execution time of the job.

**Job slowdown**: it is defined as the ratio of the response time of a job to its actual run time. It primarily occurs due to a long waiting time for execution of job.

**Economic Profit**: In utility based resource usage, it indicates profit earned by resource.

Performance evaluation metrics for application scheduling are as follows.

**Make-span**: It is calculated by subtracting actual start time of the first job of application from the actual finish time of the last job of the application.

**Schedule Length Ratio** [60]: This metric is obtained by dividing the make-span of an application by the expected time needed to execute the tasks present on the Longest Computation Path on the fastest resource.

**Scheduling time** [60]: It depends on the time-complexity of the scheduling algorithm. It indicates how much time is taken by scheduling algorithm for making decision on assignment of jobs onto resources.

**Speedup** [60]: It indicates how faster an application runs on multiple resources as compared to running it on a single fastest resource.

**The number of times best results achieved}** [60]: It indicates out of many runs of various applications, for how many runs the scheduling algorithm performed best among others.

**Flow time**: the flow time of the jobs of an application is the sum of completion time of all the jobs.

**Economic Cost**: In utility based resource usage, economic cost of an application indicates total cost incurred for executing all jobs of the application.

Of the above presented metrics, schedule length ratio is used for evaluation of only scheduling of dependent tasks application. Other metrics can be used for either independent tasks or dependent tasks applications.

*D. Decision on use of scheduling simulators or real systems for evaluation*

When a real Grid environment is not available with researchers or when it is not practically feasible and monetarily advisable to deploy a real Grid just for sake of performance evaluation of some proposed algorithm, researchers can use approach of simulation based performance evaluation. When results of experiments are needed very quickly and researchers do not have enough time to make themselves comfortable with working with real Grid environment, the simulation based experiment study is best way of evaluating proposed concepts. Furthermore, when changing configuration of real Grid environment is difficult and challenging or preparing certain scenarios are impossible to achieve, evaluation of scheduling algorithms by simulation can provide results faster and for even practically infeasible scenarios. Simulators are handy for showing proof of concepts; moreover, they can provide evaluation of proposed algorithms for scalability and efficiency performance measures. Though Grid simulators can be used for evaluation of proposed algorithms, true reliability and accuracy of the proposed algorithms can be obtained by evaluating the algorithms on real systems. If researchers have access to real Grid environment or testbed, they should evaluate proposed algorithms on the real environment.

## VI. CONCLUSIONS

This paper tried to tie discussion on various constituent software sub-systems of Grid computing based on scheduling aspect. This paper discussed concepts of scheduling in Grid computing, a complex system, involved at the lowest-level, part of operating system, to cluster system and Grid computing system. The paper also discusses importance of task scheduling in distributed environment, which is NP-Complete problem. Exact search based algorithms are not feasible and therefore researchers need to look for heuristic based algorithm. Scheduling in Grid is involved at two levels: individual site level (resource scheduling) and application broker level (application scheduling) for which this paper provided classification of various resource scheduling algorithms and application scheduling algorithms. The paper also discussed about methodology to be used for evaluating scheduling algorithms. Specifically, the paper highlighted simulation based approach and real system based approach and also discussed about when to use which one. The work also provided important performance metrics for evaluating both resource scheduling and application scheduling. In Grid computing, performance evaluation using real system takes a lot of time, efforts and requires expert skills whereas performance evaluation using simulation is relatively easy.

REFERENCES

[1] I. Foster and C. Kesselman, Eds., *The Grid 2: Blueprint for a New Computing Infrastructure*, 2nd ed., ser. The Elsevier Series in Grid Computing. Elsevier, 2003.

[2] M. Pinedo, *Scheduling: Theory, Algorithms and Systems*, 2nd ed. Prentice Hall.

[3] J. Yu and R. Buyya, "A taxonomy of workflow management systems for grid computing," *Journal of Grid Computing*, vol. 3, no. 3-4, pp. 171–200, 2005.

[4] C. S. Yeo and R. Buyya, "A taxonomy of market-based resource management systems for utility-driven cluster computing," *Softw. Pract. Exper.*, vol. 36, pp. 1381–1419, November 2006.

[5] K. Krauter, R. Buyya, and M. Maheswaran, "A taxonomy and survey of grid resource management systems for distributed computing," *Software: Practice and Experience*, vol. 32, no. 2, pp. 135–164, 2002.

[6] R. Buyya *et al.*, "High performance cluster computing: Architectures and systems (volume 1)," *Prentice Hall, Upper SaddleRiver, NJ, USA*, vol. 1, p. 999, 1999.

[7] D. W. Erwin and D. F. Snelling, "Unicore: A grid computing environment," in *Euro-Par 2001 Parallel Processing*. Springer, 2001, pp. 825–834.

[8] I. Foster and C. Kesselman, "Globus: A metacomputing infrastructure toolkit," *International Journal of High Performance Computing Applications*, vol. 11, no. 2, pp. 115–128, 1997.

[9] A. S. Grimshaw, W. A. Wulf, J. C. French, A. C. Weaver, and P. F. Reynolds, "Legion: The next logical step toward a nationwide virtual computer," Technical Report CS-94-21, University of Virginia, Tech. Rep., 1994.

[10] R. Buyya, D. Abramson, and J. Giddy, "Nimrod/g: An architecture for a resource management and seduling system in a global computational grid," in *High Performance Computing in the Asia-Pacific Region, 2000. Proceedings. The Fourth International Conference/Exhibition on*, vol. 1. IEEE, 2000, pp. 283–289.

[11] E. Deelman, D. Gannon, M. Shields, and I. Taylor, "Workflows and e-science: An overview of workflow system features and capabilities," *Future Generation Computer Systems*, vol. 25, no. 5, pp. 528–540, 2009.

[12] Montage An Astronomical Image Mosaic Engine. Last accessed on 23 July 2013. [Online]. Available: http://montage.ipac.caltech.edu/docs/gridtools.html

[13] I. Foster, C. Kesselman, and S. Tuecke, "The anatomy of the grid: Enabling scalable virtual organizations," *International journal of high performance computing applications*, vol. 15, no. 3, pp. 200–222, 2001.

[14] I. Foster, Y. Zhao, I. Raicu, and S. Lu, "Cloud computing and grid computing 360-degree compared," in *Grid Computing Environments Workshop, 2008. GCE'08*. Ieee, 2008, pp. 1–10.

[15] P. Brucker, *Scheduling Algorithms*, 3rd ed. Secaucus, NJ, USA: Springer-Verlag New York, Inc., 2001.

[16] J. Chen and C.-Y. Lee, "General multiprocessor task scheduling," *Naval Research Logistics (NRL)*, vol. 46, no. 1, pp. 57–74, 1999. [Online]. Available: http://dx.doi.org/10.1002/(SICI)1520-6750(199902)46:1<57::AID-NAV4>3.0.CO;2-H

[17] M. R. Garey and D. S. Johnson, *Computers and Intractability; A Guide to the Theory of NP-Completeness*. New York, NY, USA: W. H. Freeman & Co., 1990.

[18] R. L. Henderson, "Job scheduling under the portable batch system," in *Job scheduling strategies for parallel processing*. Springer, 1995, pp. 279–294.

[19] S. Zhou, "Lsf: Load sharing in large heterogeneous distributed systems," in *I Workshop on Cluster Computing*, 1992.

[20] M. J. Litzkow, M. Livny, and M. W. Mutka, "Condor-a hunter of idle workstations," in *Distributed Computing Systems, 1988., 8th International Conference on*. IEEE, 1988, pp. 104–111.

[21] W. Gentzsch, "Sun grid engine: Towards creating a compute power grid," in *Cluster Computing and the Grid, 2001. Proceedings. First IEEE/ACM International Symposium on*. IEEE, 2001, pp. 35–36.

[22] NQE User's Guide. Last accessed on 12 August 2013. [Online]. Available: http://docs.cray.com/books/2148_3.3/html-2148_3.3/2148_3.3-toc.html

[23] Maui Scheduler. Last accessed on 12 August 2013. [Online]. Available: http://www.adaptivecomputing.com/products/open-source/maui/

[24] S. Kannan, M. Roberts, P. Mayes, D. Brelsford, and J. F. Skovira, "Workload management with loadleveler," *IBM Redbooks*, vol. 2, p. 2, 2001.

[25] M. L. Massie, B. N. Chun, and D. E. Culler, "The ganglia distributed monitoring system: design, implementation, and experience," *Parallel Computing*, vol. 30, no. 7, pp. 817–840, 2004.

[26] R. Wolski, N. T. Spring, and J. Hayes, "The network weather service: a distributed resource performance forecasting service for metacomputing," *Future Generation Computer Systems*, vol. 15, no. 5, pp. 757–768, 1999.

[27] R. Wolski, "Dynamically forecasting network performance using the network weather service," *Cluster Computing*, vol. 1, no. 1, pp. 119–132, 1998.

[28] F. Berman, R. Wolski, H. Casanova, W. Cirne, H. Dail, M. Faerman, S. Figueira, J. Hayes, G. Obertelli, J. Schopf *et al.*, "Adaptive computing on the grid using apples," *Parallel and Distributed Systems, IEEE Transactions on*, vol. 14, no. 4, pp. 369–382, 2003.

[29] K. Seymour, A. YarKhan, S. Agrawal, and J. Dongarra, "Netsolve: Grid enabling scientific computing environments," *Advances in Parallel Computing*, vol. 14, pp. 33–51, 2005.

[30] N. Furmento, W. Lee, A. Mayer, S. Newhouse, and J. Darlington, "Iceni: an open grid service architecture implemented with jini," in *Proceedings of the 2002 ACM/IEEE conference on Supercomputing*. IEEE Computer Society Press, 2002, pp. 1–10.

[31] I. Taylor, M. Shields, and I. Wang, "Resource management for the triana peer-to-peer services," in *Grid Resource Management*. Springer, 2004, pp. 451–462.

[32] K. Amin, G. Von Laszewski, M. Hategan, N. J. Zaluzec, S. Hampton, and A. Rossi, "Gridant: A client-controllable grid workflow system," in *System Sciences, 2004. Proceedings of the 37th Annual Hawaii International Conference on*. IEEE, 2004, pp. 10–pp.

[33] G. von Laszewski and M. Hategan, "Workflow concepts of the java cog kit," *Journal of Grid Computing*, vol. 3, no. 3-4, pp. 239–258, 2005.

[34] T. Fahringer, A. Jugravu, S. Pllana, R. Prodan, C. Seragiotto, and H.-L. Truong, "Askalon: a tool set for cluster and grid computing," *Concurrency and Computation: Practice and Experience*, vol. 17, no. 2-4, pp. 143–169, 2005.

[35] S. McGough, L. Young, A. Afzal, S. Newhouse, and J. Darlington, "Workflow enactment in iceni," in *UK e-Science All Hands Meeting*, 2004, pp. 894–900.

[36] J. Frey, "Condor dagman: Handling inter-job dependencies," *University of Wisconsin, Dept. of Computer Science, Tech. Rep*, 2002.

[37] D. Hull, K. Wolstencroft, R. Stevens, C. Goble, M. R. Pocock, P. Li, and T. Oinn, "Taverna: a tool for building and running workflows of services," *Nucleic acids research*, vol. 34, no. suppl 2, pp. W729–W732, 2006.

[38] F. Berman, A. Chien, K. Cooper, J. Dongarra, I. Foster, D. Gannon, L. Johnsson, K. Kennedy, C. Kesselman, J. Mellor-Crumme *et al.*, "The grads project: Software support for high-level grid application development," *International Journal of High Performance Computing Applications*, vol. 15, no. 4, pp. 327–344, 2001.

[39] J. Cao, S. A. Jarvis, S. Saini, and G. R. Nudd, "Gridflow: Workflow management for grid computing," in *Cluster Computing and the Grid, 2003. Proceedings. CCGrid 2003. 3rd IEEE/ACM International Symposium on*. IEEE, 2003, pp. 198–205.

[40] R. Buyya and S. Venugopal, "The gridbus toolkit for service oriented grid and utility computing: An overview and status report," in *Grid Economics and Business Models, 2004. GECON 2004. 1st IEEE International Workshop on*. IEEE, 2004, pp. 19–66.

[41] E. Deelman, G. Singh, M.-H. Su, J. Blythe, Y. Gil, C. Kesselman, G. Mehta, K. Vahi, G. B. Berriman, J. Good, A. Laity, J. C. Jacob, and D. S. Katz, "Pegasus: A framework for mapping complex scientific workflows onto distributed systems," *Sci. Program.*, vol. 13, pp. 219–237, July 2005. [Online]. Available: http://dl.acm.org/citation.cfm?id=1239649.1239653

[42] R. Jain, *The art of computer systems performance analysis*. John Wiley & Sons, 2008.

[43] H. Casanova, "Simgrid: A toolkit for the simulation of application scheduling," in *Cluster Computing and the Grid, 2001. Proceedings. First IEEE/ACM International Symposium on*. IEEE, 2001, pp. 430–437.

[44] R. Buyya and M. Murshed, "Gridsim: a toolkit for the modeling and simulation of distributed resource management and scheduling for grid computing," *Concurrency and Computation: Practice and Experience*, vol. 14, no. 13-15, pp. 1175–1220, 2002. [Online]. Available: http://dx.doi.org/10.1002/cpe.710

[45] W. H. Bell, D. G. Cameron, L. Capozza, A. P. Millar, K. Stockinger, and F. Zini, "Optorsim - a grid simulator for studying dynamic data replication strategies," *International Journal of High Performance Computing Applications*, vol. 17, no. 4, pp. 403–416, 2003.

[46] A. Chervenak, I. Foster, C. Kesselman, C. Salisbury, and S. Tuecke, "The data grid: Towards an architecture for the distributed management and analysis of large scientific datasets," *Journal of network and computer applications*, vol. 23, no. 3, pp. 187–200, 2000.

[47] D. Klusácek and H. Rudová, "Alea 2: job scheduling simulator," in *Proceedings of the 3rd International ICST Conference on Simulation Tools and Techniques*, ser. SIMUTools '10. ICST, Brussels, Belgium, Belgium: ICST (Institute for Computer Sciences, Social-Informatics and Telecommunications Engineering), 2010, pp. 61:1–61:10. [Online]. Available: http://dx.doi.org/10.4108/ICST.SIMUTOOLS2010.8722

[48] K. Kurowski, J. Nabrzyski, A. Oleksiak, and J. Weglarz, "Grid scheduling simulations with gssim," in *Parallel and Distributed Systems, 2007 International Conference on*, vol. 2. IEEE, 2007, pp. 1–8.

[49] Y. Huang, A. Brocco, M. Courant, B. Hirsbrunner, and P. Kuonen, "Magate simulator: a simulation environment for a decentralized grid scheduler," in *Advanced Parallel Processing Technologies*. Springer, 2009, pp. 273–287.

[50] W. Chen and E. Deelman, "Workflowsim: A toolkit for simulating scientific workflows in distributed environments," in *E-Science (e-Science), 2012 IEEE 8th International Conference on*. IEEE, 2012, pp. 1–8.

[51] K. Kurowski, A. Oleksiak, W. Piatek, and J. Weglarz, "Hierarchical scheduling strategies for parallel tasks and advance reservations in grids," *Journal of Scheduling*, pp. 1–20, 2011.

[52] Computing with HTCondor™. Last accessed on 23 July 2013. [Online]. Available: http://research.cs.wisc.edu/htcondor/

[53] H. Chen and M. Maheswaran, "Distributed dynamic scheduling of composite tasks on grid computing systems," in *Parallel and Distributed Processing Symposium., Proceedings International, IPDPS 2002, Abstracts and CD-ROM*, 2002, pp. 88 – 97.

[54] F. Dong and S. G. Akl, "Scheduling algorithms for grid computing: State of the art and open problems," *School of Computing, Queen's University, Kingston, Ontario*, 2006.

[55] J. Yu, R. Buyya, and K. Ramamohanarao, "Workflow scheduling algorithms for grid computing," in *Metaheuristics for scheduling in distributed computing environments*. Springer, 2008, pp. 173–214.

[56] M. Maheswaran, S. Ali, H. Siegal, D. Hensgen, and R. F. Freund, "Dynamic matching and scheduling of a class of independent tasks onto heterogeneous computing systems," in *Heterogeneous Computing Workshop, 1999.(HCW'99) Proceedings. Eighth*. IEEE, 1999, pp. 30–44.

[57] S. K. Garg, R. Buyya, and H. J. Siegel, "Scheduling parallel applications on utility grids: time and cost trade-off management," in *Proceedings of the Thirty-Second Australasian Conference on Computer Science - Volume 91*, ser. ACSC '09. Darlinghurst, Australia, Australia: Australian Computer Society, Inc., 2009, pp. 151–160.

[58] J. Yu, R. Buyya, and C. K. Tham, "Cost-based scheduling of scientific workflow application on utility grids," in *Proceedings of the First International Conference on e-Science and Grid Computing*. Washington, DC, USA: IEEE Computer Society, 2005, pp. 140–147.

[59] "Benchmarking the task graph scheduling algorithms," in *Proceedings of the 12th. International Parallel Processing Symposium on International Parallel Processing Symposium*, ser. IPPS '98. Washington, DC, USA: IEEE Computer Society, 1998, pp. 531–.

[60] H. Topcuouglu, S. Hariri, and M.-y. Wu, "Performance-effective and low-complexity task scheduling for heterogeneous computing," *IEEE Trans. Parallel Distrib. Syst.*, vol. 13, pp. 260–274, March 2002.